\documentclass[epj,draft]{svjour}
\input{epsf}
\begin{document}
\title{Two interacting particles in a disordered chain IV: \\ 
Scaling of level curvatures}
\author{Andr\'e Wobst \and Dietmar Weinmann\thanks{
\email{Dietmar.Weinmann@Physik.Uni-Augsburg.DE}}}
\institute{Institut f\"ur Physik, Universit\"at Augsburg, 86135 Augsburg, 
           Germany }
\date{\today}
\abstract{
The curvatures of two-particle energy levels with respect to the enclosed 
magnetic flux in mesoscopic disordered rings are investigated numerically. We 
find that the typical value of the curvatures is increased by interactions in 
the localised regime and decreased in the metallic regime. This confirms a 
prediction by Akkermans and Pichard (Eur.\ Phys.\ J.\ B {\bf 1}, 223 (1998)). 
The interaction-induced changes of the typical curvatures at different energies
and disorder strengths exhibit one-parameter scaling with a conductance-like 
single parameter. This suggests that interactions could influence the 
conductance of mesoscopic systems similarly. 
}
\maketitle

\section{Introduction}

For disordered mesoscopic systems, the curvatures of one-particle energy levels
as a function of an external parameter like the magnetic flux enclosed by a 
system with ring geometry have been studied intensively since Edwards and 
Thouless \cite{edwards,thouless} proposed that the typical value of the 
curvatures should be closely related to the non-interacting conductance.
Intuitively, it is clear that level curvatures and the conductance should be 
related. Since the effect of an enclosed flux is equivalent to generalised 
boundary conditions, the curvatures measure the sensitivity of the energy 
levels with respect to the boundary conditions. Obviously, eigenstates which 
are localised are affected much less by the boundaries than extended states and
the level curvatures are strongly reduced by localisation. The argument given 
by Edwards and Thouless is based on the occurrence of momentum matrix elements 
in a perturbative expansion of the flux dependence of the eigenenergies as well
as in the Kubo formula for the conductance. It follows that the dimensionless 
conductance $g_1$ should be proportional to the typical curvature $\tilde{c}$ 
divided by the mean level spacing $\Delta_1$. Several analytical studies 
support this Thouless relation, and it has also been verified numerically 
within the Anderson model \cite{braun}. The mean of the absolute value of the 
curvatures is indeed proportional to a properly defined conductance in the 
ballistic and the diffusive regime whereas in the localised regime, the mean of
the logarithms of curvatures and conductance were found to be proportional. In 
addition, it is known that the energy level statistics at fixed flux also 
contains an energy scale (the so-called Thouless-energy), which is proportional
to the typical one-particle curvature and thus to the conductance. 

Furthermore, it was shown that not only the mean value of the curvatures but 
also their distribution depends on the conductance. In the diffusive regime 
\cite{zyczkowski,canali,braun-mont}, the distribution is very close to that 
obtained from random matrix theory for the Gaussian orthogonal ensemble (GOE)
\cite{oppen-curve} while a log-normal distribution is approached in the 
localised regime \cite{braun,zyczkowski,canali,titov}. 

Although the relation between curvatures and conductance has been intensively 
studied in the non-interacting case, very little is known about the relevance 
of the curvatures of many-body levels in disordered systems with 
electron-electron interactions.  Only persistent currents which are related to 
the first derivative of the many-body ground state energy with respect to flux 
have been studied in detail in the presence of interactions 
\cite{eckern,moriond}. Several theories predict an enhancement of persistent 
currents due to interactions. Recently, very strong enhancement has been found 
for spinless fermions in individual one-dimensional rings at large disorder 
strength \cite{schmitteckert}. However, it is not yet clear whether this can 
fully account for the experimental findings in the diffusive regime which 
suggest values up to two orders of magnitude larger than the predictions of 
theories neglecting interactions.

The interplay of disorder and interaction is a subject of very intense research
activities \cite{moriond}. While perturbative approaches can be used for weak 
disorder or weak interaction strength in the thermodynamic limit, a relatively 
new approach considers a small number of particles with both, strong disorder 
and strong interaction. It has been predicted \cite{shep,imry} and confirmed 
numerically \cite{fmgpw,wmgpf,oppen1} that a short-range interaction enhances 
the localisation length of certain two-particle states in disordered systems. 
However, energy dependent studies for quasi-particles have shown that the 
effect is suppressed at small excitation energies \cite{oppen2}.

Since this enhancement effect can be connected to the two-particle energy level
statistics \cite{wepi}, the latter has been investigated in detail. In the
non-interacting case, two-particle energies are sums of two one-particle 
energies, and therefore the two-particle level spacing distribution is always 
close to the uncorrelated Poisson distribution, even in metallic systems where 
the one-body level statistics is close to the universal Wigner-Dyson statistics
for the GOE. In contrast, the presence of interactions leads to spectral 
correlations with interesting properties. For on-site interactions on a 
one-dimensional lattice, there exists a duality transformation which maps the 
behaviour at weak interaction to the one at very strong interaction 
\cite{tip2,si}. Therefore, the spectrum at infinite interaction is 
uncorrelated like in the non-interacting case. Maximal correlations occur for 
intermediate interaction strength. In the band centre, these correlations are 
very close to the critical ones found in several other systems exhibiting weak 
chaos like the Anderson model in three dimensions at the critical point 
\cite{braun1}. The appearance of the critical statistics for two interacting 
particles \cite{tip2} is accompanied by multi-fractal wavefunctions. This may 
be caused by the multi-fractal structure of the interaction matrix itself 
\cite{tip1}.    

While the spectral correlations seem to be closely related to the 
interaction-induced enhancement of the localisation length in the localised 
regime \cite{wepi}, it is less obvious whether and how they can be used to 
predict observable quantities like the conductance of metallic samples. It is 
therefore necessary to consider properties of the two-particle system which
are more directly connected to the mobility of the particles. An important step
in this direction is a numerical study of the real-time development of 
two-particle wave-packets \cite{tip3}, where the time-dependence of the 
spreading of two particles released at adjacent sites is observed, yielding 
useful informations on the influence of the interactions. While they reduce the
speed of the ballistic spreading of the wave-packet at very short times, they 
lead on the other hand to an extension beyond the one-particle localisation 
length $L_1$ for intermediate times and the larger two-particle localisation 
length $L_2$ is reached sub-diffusively after very long time \cite{tip3}. While
long times yield informations about the localised regime, the short time 
dynamics should be connected to the behaviour of metallic samples. 

The level curvatures with respect to an external parameter like an enclosed 
flux represent a static quantity which is directly related to the mobility of 
the particles in metallic and in insulating samples. Very recently, 
Akkermans and Pichard \cite{ap} have studied the connection between spectral 
correlations and level curvatures for two interacting particles. They predicted
that the curvatures should be enhanced by interactions in localised and reduced
in metallic samples. However, the analytic arguments used to obtain this very 
interesting result contain crude approximations and it is not obvious how to 
extract details like the curvature distributions or the energy dependence and 
information about the crossover between the two regimes. This has motivated us 
to perform a numerical study of the two-particle level curvatures in a 
one-dimensional ring. We have unambiguously confirmed the prediction mentioned
above \cite{ap}. Furthermore, our data suggest that the effect of the 
interaction on curvatures exhibits one-parameter scaling. The scaling parameter
is closely related to the non-interacting conductance. Many of the observed 
features are closely related to the spectral statistics and the dynamics of 
wave-packets studied in the preceding papers \cite{tip2,tip1,tip3} of this 
series on two interacting particles in a disordered chain.

After introducing the system in Section \ref{system}, we explain in Section 
\ref{method} our numerical techniques and discuss different averaging 
procedures. The results for the effect of the interaction on the typical level 
curvatures are presented in Section \ref{curve-mean}. Details of the curvature 
distribution are shown in Section \ref{curve-distribution} before we present 
our conclusions.

\section{Disordered ring}\label{system}

The Hamiltonian for two interacting particles can be written in the form
\begin{equation}\label{hamiltonian}
H=H_1\otimes {\bf 1} + {\bf 1} \otimes H_1+H_{\rm int}\, .
\end{equation}
$H_1$ describes one particle in a disordered chain with $L$ sites and reads
\begin{equation}
H_1=\sum_{n=0}^{L-1}\Big(-t|n+1\rangle\langle n|-t^*|n\rangle
\langle n+1|+V_n|n\rangle\langle n|\Big) \, ,
\end{equation}
where $|n\rangle$ is a Wannier function localised at site $n$. Periodic
boundary conditions close the chain to a ring, implying 
$|n\rangle \equiv |n+L\rangle$. The disorder is modeled by random on-site 
energies $V_n$ which are distributed uniformly inside the interval $[-W:W]$. 
The nearest neighbor hopping matrix elements 
\begin{equation}
t=t_0\exp(2\pi{\rm i}\phi/L)\, ,
\end{equation}
whose absolute value $t_0$ is set to unity and defines the energy scale, 
contain the dimensionless flux $\phi=\varphi/\varphi_0$ which is given by the 
magnetic flux $\varphi$ threading the ring in units of the flux quantum 
$\varphi_0=hc/e$. The magnetic flux affects only the phase of the hopping 
matrix elements $t$ and can be transformed by a gauge transformation into a 
generalised periodic boundary condition with a phase shift of $2\pi\phi$. 
Therefore, a variation of the magnetic flux is equivalent to a change of the 
boundary condition. For the one-particle case the curvatures 
\begin{equation}
c_\alpha=\left. \frac{\partial^2 \epsilon_\alpha}{\partial\phi^2}
\right|_{\phi=0}
\end{equation}
of the eigenvalues $\epsilon_\alpha $ of $H_1$ are directly related to the 
conductance by means of the Thouless relation. 

For simplicity, we use a Hubbard-like on-site interaction with strength $U$
between the two particles described by
\begin{equation}
H_{\rm int}=\sum_{n=0}^{L-1}U|nn\rangle\langle nn|\, .
\end{equation}
Here, $|mn\rangle$ means $|m\rangle|n\rangle$ and restricting ourselves to the 
case of two fermions with opposite spins, the dimension of the Hilbert space is
$M=L(L+1)/2$. A complete basis for the symmetric configuration space part of 
the states is given by 
\begin{equation} \label{basis}
\begin{array}{rcl}|mn\rangle_s&=&\frac{1}{\sqrt{2}}(|mn\rangle+|nm\rangle)
\quad\mbox{for}\quad m\ne n \\ |nn\rangle_s&=&|nn\rangle \, . \end{array}
\end{equation} 
 
\section{Numerical techniques}\label{method}

\subsection{Computation of level curvatures}

We write the Hamiltonian (\ref{hamiltonian}) in the symmetric two-particle 
basis (\ref{basis}). Since the resulting $M\times M$ matrix is very sparse, we 
use the Lanczos algorithm to numerically calculate the full set of $M$
two-particle eigenenergies.

The absolute values of the curvatures of different levels (even within the same
disorder realisation) usually differ by many orders of magnitude, making it 
difficult to obtain reliable numerical results for small as well as for large 
curvatures. Therefore we calculate the two-particle eigenenergies at several 
values of the magnetic flux between $\phi=0$ and $\phi\sim 10^{-3}$. For each 
of the $M$ eigenvalues, we first estimate the fourth derivative with respect to
the flux. Then, we perform a least square fit with a parabola to the calculated
data. The weights of the data points at different flux values are adjusted 
individually for each of the eigenvalues as a function of the estimates for the
fourth derivative and the typical inaccuracy of the numerically calculated 
eigenvalues \cite{andre}.

Another possibility is to use the perturbative expression for the eigenenergies
in second order in the magnetic flux. This allows to calculate the curvatures 
directly and without the errors arising from the numerical calculation of 
derivatives. However, this requires the eigenenergies and in addition all of 
the eigenvectors of the Hamiltonian matrix at $\phi=0$. Since this considerably
increases the computation time and required memory, we have not used this 
approach for the large matrices occurring in the interacting problem. However, 
in the one-particle case such a calculation is easily possible for the ring 
sizes we consider and we have used this to adjust our general method and to 
verify its precision by comparing two-particle results without interaction. We 
believe that this yields a good estimate for the precision of the curvatures 
calculated at arbitrary interaction strength. 

\subsection{Averaging procedures}

\begin{figure}[tb]
\centerline{\epsfxsize=8 cm \epsffile{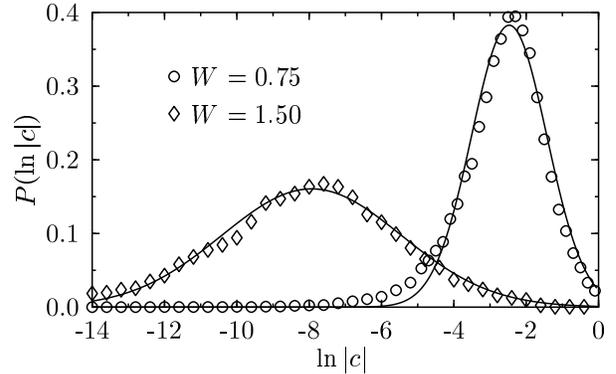}}
\caption{Typical distributions $P(\ln|c|)$ of curvatures for two-particle 
levels in the energy interval [-2.5 : -1] at interaction strength $U=0.5$ and 
ring size $L=100$. The solid lines are Gaussian fits.}\label{pcsample}
\end{figure}

Since we want to discuss typical values of the curvatures as a function of 
disorder and interaction strength, the averaging technique must be chosen 
carefully. Besides the mean of the absolute values and the mean of the 
logarithm of the absolute values, we have used another way to extract a typical
curvature, exploiting the fact that there is a peak in the distribution 
$P(\ln|c|)$ as shown in Figure~\ref{pcsample} and a fit of a log-normal 
distribution is usually rather good. Such a log-normal curvature distribution 
has been observed numerically in the insulating regime of the Anderson model 
\cite{braun,zyczkowski,canali} and was derived analytically for one-dimensional
disordered rings \cite{titov}. The peak width grows with the disorder strength 
and the position of the maximum decreases. This is reminiscent of the behaviour
of one particle in a two-dimensional disordered system in the localised regime 
where the dimensionless conductance $g_1$ is log-normally distributed with 
typical value $\langle\ln g_1\rangle\propto -L/L_1$ with a variance which is 
given by the mean value $-\langle\ln g_1\rangle$ \cite{pich}. Our data for the 
curvatures are rather well approximated by the estimate 
$\langle\ln |c|\rangle=\ln(4\pi^2\Delta_1)-L/L_1$, where $\Delta_1$ denotes the
average one-particle level spacing and also show a variance close to 
$-\langle\ln |c|\rangle$. We found that the most reproducible results for 
typical curvatures are obtained by fitting the distribution of curvatures 
(rather than computing averages) because this minimises the influence of the 
statistically less accurate points in the tails of the distribution. We always 
used a log-normal fit for this purpose as shown in Figure~\ref{pcsample}, 
yielding numerically reliable results for the average $\langle \ln|c| \rangle$ 
and introduce $\tilde{c}=\exp(\langle \ln|c| \rangle)$ as a typical curvature. 
Additionally, we qualitatively confirmed our results using the numerically less
stable direct averages of $\ln|c|$ and $|c|$.

\section{Typical curvatures}\label{curve-mean}

\subsection{Energy dependence without interaction}

The energy dependence of the typical two-particle level curvature is shown in 
Figure~\ref{cvse}. In order to understand the non-interacting curve (full 
symbols), let us first recall the energy dependence of the curvatures of 
one-particle levels in one-dimensional rings. In the average over disorder 
realisations the curvatures are small at the band edges but several orders of 
magnitude larger in the band centre. Their sign is alternating with the number 
of the state. In the non-interacting two-particle case different regions can be
observed. The edges at large absolute energy close to the band edges at 
$|E|=2(2t_0+W)$ arise from one-particle states where both of the particles are 
close to the same edge of the one-particle band where their typical curvature 
is very small. Therefore the two-particle level curvatures are small at the 
band edges. To obtain a state in the centre of the two-particle band, there are
several possibilities to combine two one-particle states such that their total 
energy is close to $E=0$. They can be chosen at the two opposite edges of the 
one-particle spectrum, both having small one-particle curvature and therefore 
yielding a small two-particle curvature. The other extreme is represented by 
both particles occupying states close to the middle of the one-particle band. 
Then the individual curvatures are much larger but may have different sign. The
typical two-level curvature in this region is given by an average over all of 
these possibilities. For intermediate energies $|E|\approx 2$ most of the 
two-particle states are built from one particle in the centre and the second 
particle close to an edge of the one-particle band. Since there are no 
contributions from states composed by two particles both located close to an 
edge of the one-particle band, and furthermore cancellations due to different 
signs are less important because of different absolute values of the individual
curvatures, the typical two-particle level curvatures in these energy ranges 
are larger than in the centre of the band where a mixing of many different 
regimes affects the results.

\subsection{Influence of the interaction on curvatures}

\begin{figure}[tb]
\centerline{\epsfxsize=8 cm \epsffile{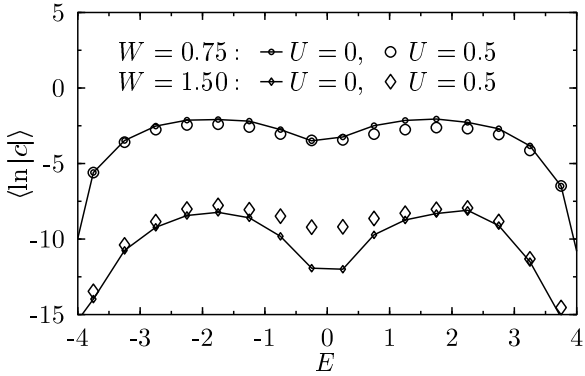}}
\caption{The energy dependence of typical curvatures for different disorder 
strengths $W$ and $L=100$. $\langle\ln|c|\rangle$ is plotted for the cases
without (full symbols, connected by lines) and with interaction $U=0.5$ (open 
symbols).}\label{cvse}
\end{figure}

Interactions have a significant influence on the typical two-particle level 
curvatures. This can be seen by comparing the open symbols in Figure~\ref{cvse}
(which are for $U=0.5$) to the non-interacting case given by the full symbols. 
The influence of the interaction is shown for disorder strengths $W=0.75$ and 
$W=1.5$. The main observation is that while at low disorder interactions 
decrease the curvatures, the latter are increased at higher disorder when the 
one-particle states are localised. This is a direct confirmation of the 
prediction by Akkermans and Pichard \cite{ap}. A closer look at the data 
reveals a strong dependence of the effect on the energy regime with a tendency 
to remove the dip in the curves around $E=0$.

Although at first glance positive and negative energies seem to show the same 
behaviour in Figure~\ref{cvse}, there are small but significant differences. 
However, the data obey the symmetry relation $\tilde{c}(U,E)=\tilde{c}(-U,-E)$
because the ensemble of Hamiltonians remains unchanged when both, the sign of 
the energy and the sign of the interaction are changed. Thus, the data for 
positive and negative energy should be the same only at $U=0$, and deviations
may occur at finite interaction strength. We have checked that the small 
difference in our results between positive and negative energy is indeed 
inverted when the sign of the interaction $U$ is changed, even though the
averages have been performed over a limited number of samples. Therefore, it is
sufficient to consider only interactions of a definite sign and we restrict 
ourselves to the case of positive $U$.

\subsection{One parameter scaling}

\begin{figure}[tb]
\centerline{\epsfxsize=8 cm \epsffile{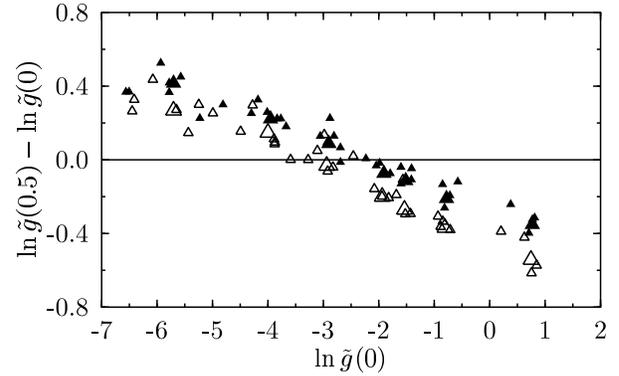}}
\caption{The difference of the logarithms of the dimensionless curvatures 
$\tilde{g}(U)$ in rings of size $L=100$ at interaction strength $U=0.5$ and at 
$U=0$ versus the logarithm of the conductance-like $\tilde{g}(0)$. Full and 
open triangles are for negative and positive energy, respectively. Large 
triangles represent averages over the energy ranges $[-2.5:-1]$ and $[1:2.5]$, 
small triangles are (less precise) averages over narrower energy ranges of 
width 0.5 within the intervals $[-4:-1]$ and $[1:4]$. The data for disorder 
strengths between $W=0.75$ and $W= 1.5$ and the different energies are 
consistent with two straight lines, one for each sign of the energy.}
\label{cvsc}
\end{figure}

Since the influence of the interaction depends on the mobility of the particles
(increase for localised, decrease for metallic samples), it is tempting to
define a dimensionless curvature
\begin{equation}
\tilde{g}(U)=\tilde{c}/\Delta_1
\end{equation}
where $\tilde{c}$ is the typical curvature at interaction $U$ and
$\Delta_1\approx 2(2t_0+W)/L$ is the mean one-particle level spacing, and to 
plot the interaction-induced change of curvatures not as a function of energy 
and disorder, but over the conductance-like parameter $\tilde{g}(0)$. The 
latter is given by the typical non-interacting two-particle curvature
$\tilde{c}$ at two-particle energy $E$ in units of $\Delta_1$ and therefore
represents an average over the one-particle Thouless-conductances $g_1$ at the 
different contributing one-particle energies.

In Figure~\ref{cvsc}, the difference between the logarithms of the typical 
absolute values of the dimensionless curvatures with interaction $U=0.5$ and 
without interaction are plotted versus the logarithm of the conductance-like 
parameter $\tilde{g}(0)$ for different energy ranges and disorder strengths 
$W$. The large triangles are for the energy intervals $[-2.5:-1]$ and 
$[1:2.5]$, where the energy dependence of the typical value but also of the 
shape of the distribution $P(\ln|c|)$ is very small so that statistically 
reliable and precise data can be obtained by averaging not only over many 
disorder realisations, but also over a relatively broad energy range. On the 
other hand, we have omitted the centre of the band in this plot because for 
strong disorder the distribution $P(\ln|c|)$ shows no significant peak and the 
error estimates become considerably larger there. 

For a given sign of the energy, the data plotted in Figure~\ref{cvsc} suggest 
that the interaction-induced change of the curvatures depends not on the energy
and the disorder separately, but on only one parameter which is indeed the 
conductance-like $\tilde{g}(0)$. This one-parameter scaling is universal within
a given sign of the energy. The scaling curves for positive and negative energy
differ by a shift of the curvature scale. No change of the scaling curves is 
observed when the ground state energy is approached. This means that the effect
is relevant also for energetically low-lying excitations dominating the 
transport properties at low temperatures. That interactions indeed reduce or 
enhance the low-temperature conductance depending on the ratio between disorder
and kinetic energy was shown by direct numerical investigations of the 
energetically lowest many-body states in small disordered two-dimensional 
systems \cite{vojta}.  

The results shown in Figure~\ref{cvsc} clearly demonstrate the existence of a 
critical $\tilde{g}(0)=\tilde{g}_{\rm{crit}}$ below which the interaction
enhances two-particle level curvatures, and above which the interaction tends
to decrease the typical curvatures. This critical value is
$\ln\tilde{g}_{\rm crit}\approx -3$ for positive energies and
$\ln\tilde{g}_{\rm crit}\approx -2.2$ for negative energies\footnote{The values
depend slightly on the averaging procedure. By regarding other averages, we 
have checked that this does not affect the qualitative results.}. Since the 
level curvature without interaction is related to the conductance of the 
system, this clearly confirms again that the sign of the interaction-induced 
change indeed depends on the transport properties of the non-interacting system
as predicted \cite{ap}. With an estimate of the typical one-particle curvature 
one finds
\begin{equation}
\tilde{g}(0) \approx (2\pi)^2 \exp(-L/L_1)
\end{equation} 
and the critical conductances above correspond to localisation lengths $L_1$ 
which are about 5 to 6 times smaller than the circumference of the ring. 
Lower conductances $\tilde{g}(0)<\tilde{g}_{\rm crit}$ correspond to localised
states and  the enhancement of the curvatures is a consequence of the
enhancement of the two-particle localisation length by the interactions
proposed by Shepelyansky \cite{shep}. Larger conductances 
$\tilde{g}(0)>\tilde{g}_{\rm crit}$ can be interpreted as indicating more or 
less extended one-particle states for which the transport is suppressed by the 
interactions.     

\begin{figure}[tb]
\centerline{\epsfxsize=8 cm \epsffile{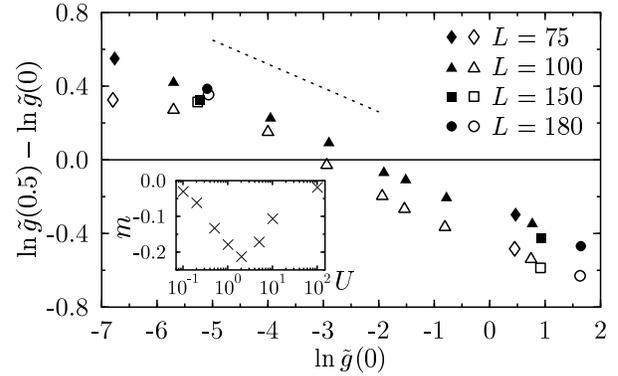}}
\caption{The difference of the logarithms of the dimensionless curvatures
$\tilde{g}(U)$ at interaction strength $U=0.5$ and $U=0$ versus the logarithm 
of $\tilde{g}(0)$. Full symbols represent the energy range $[-2.5:-1]$, open 
symbols $[1:2.5]$. Different ring sizes $L=75$ with $W=0.9$ and $W=1.8$, 
$L=100$ with the same parameters as in Figure~\ref{cvsc}, $L=150$ with $W=0.6$ 
and $W=1.2$, $L=180$ with $W=0.45$ and $W=1.1$. The dotted line has the same 
slope $m\approx -0.13$ as the scaling curves. The inset shows the dependence 
$m(U)$ of this slope on the interaction strength $U$.}
\label{cvscdiffsize}
\end{figure}

The one-parameter scaling of the change of curvatures with the non-interacting 
conductance holds also when the system size is varied as shown in 
Figure~\ref{cvscdiffsize}, where data for different ring sizes between $L=75$ 
and $L=180$ are shown. Within the numerical errors, the data lie on the scaling
curve which corresponds to their sign of energy\footnote{The small deviations 
observed can be explained as follows. For each system size the disorder average
has been done over the same set of seeds for the random number generator 
yielding nice scaling curves in spite of the limited number of different 
realisations. When comparing different system sizes however it is not possible 
to use the same realisations and the fluctuations of the data from one set of 
samples to another one (see Figure~\ref{cvsu} below) reduces the relative 
accuracy of the data points corresponding to different system sizes.}. In the 
logarithmic representation shown in Figure~\ref{cvscdiffsize} the scaling 
curves are very close to straight lines. The slopes $m(U)$ vary between $m=0$ 
(at $U=0$ and $U=\infty$) and the minimal value $m\approx-0.2$ at intermediate 
interaction strength. Such a scaling law corresponds to a power law dependence 
\begin{equation}
\frac{\tilde{g}(U)}{\tilde{g}(0)}\approx
\left(\frac{\tilde{g}(0)}{\tilde{g}_{\rm crit}}\right)^{m(U)}\, .
\end{equation}
We have observed interaction-induced changes of the dimensionless curvatures 
larger than a factor of 2. Thus the effect of the interaction can be rather 
important.  

Our observation is closely related to what is found for the dynamics of 
two-particle wave-packets \cite{tip3}. The interaction leads to a slower 
spreading at short times, before $L_1$ is reached. For these short times, only
a part of the system smaller than $L_1$ is explored by the particles and their 
reduced mobility corresponds to the decrease of the curvarure in the metallic 
regime when $\tilde{g}(0)$ is large. On the other hand, above a characteristic
time $t_1$, a slow interaction assisted propagation continues to increase the 
size of the wave-packet beyond the non-interacting saturation. This is 
reminiscent of the increase of curvatures in insulating samples when 
$\tilde{g}$ is small. 

\subsection{Duality between weak and strong interaction}

\begin{figure}[tb]
\centerline{\epsfxsize=8 cm \epsffile{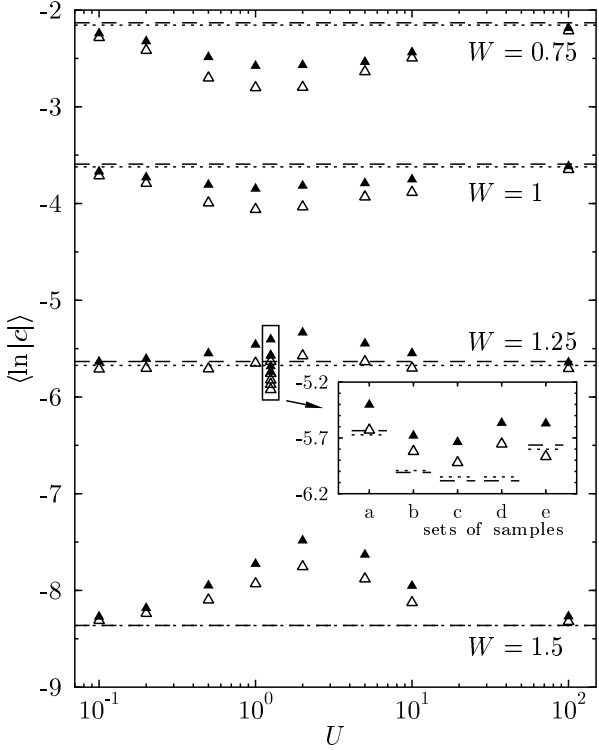}}
\caption{The typical curvatures as a function of the interaction strength $U$
and rings of size $L=100$. Symbols show the dependence of 
$\langle\ln|c|\rangle$ on the interaction, while lines indicate the value at 
$U=0$. Full symbols and dashed lines represent the energy range $[-2.5:-1]$, 
open symbols and dotted lines are for $[1:2.5]$. The inset illustrates the size
of the fluctuations due to different disorder realisations.}\label{cvsu}
\end{figure}

For $U=0$, the eigenstates of the two-particle Hamiltonian are symmetrized 
products of two one-particle states. Due to the local character of the 
interaction, this is also the case at infinite interaction $U$ \cite{tip2,si}, 
with the only complication that one has to use one-particle eigenstates of the 
system at $\phi=1/2$ in order to obtain the two-particle eigenstates at 
$\phi=0$. Therefore the typical curvatures are the same in the two limits
\footnote{This is true except for the 'molecular states', where both particles 
are localised on the same site. For large systems their number is 
parametrically smaller than the total number of states and their influence can 
be neglected.}. Due to a duality transformation for the interacting 
Hamiltonian, the effective interaction matrix elements between the eigenstates 
at $U=0$ and $U=\infty$ have the same statistical properties at small $U$ and 
small $\sqrt{24}\, t_0^2/U$, respectively. Thus, the spectral correlations are 
the same in the vicinity of the two limits \cite{tip2}. 

The behaviour of the slope of the scaling curves $m(U)$ (inset of 
Figure~\ref{cvscdiffsize}) shows that the consequences of this duality 
transformation can also be observed for the curvatures. The full interaction 
dependence of typical curvatures is presented in Figure~\ref{cvsu}. As expected
from the duality, the influence of the interaction increases up to a value 
which is of the order $U\sim t_0=1$ and returns to the non-interacting value 
when the interaction is increased further. 

As shown in the inset of Figure~\ref{cvsu}, averages over several different 
sets of 20 disorder realisations fluctuate significantly (the data at other 
interaction strength and disorder were calculated with the random seeds from 
set ``a''), and it is difficult to find accurate absolute values for the 
typical curvatures. However, the interaction dependence is similar for all sets
of samples.

\begin{figure}[tb]
\centerline{\epsfxsize=8 cm \epsffile{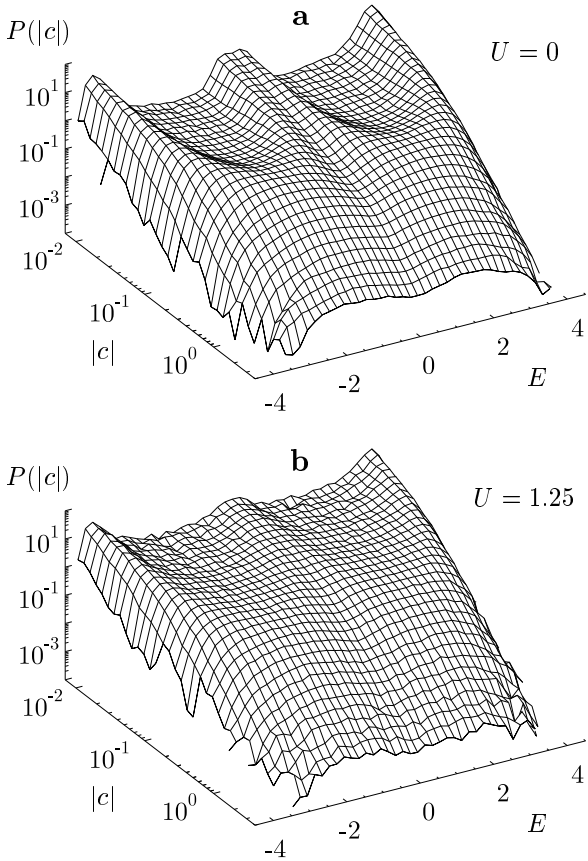}}
\caption{The distribution $P(|c|)$ for rings of size $L=100$ at disorder 
strength $W=0.5$ (a) without interaction and (b) with interaction $U=1.25$. The
average has been performed over 1000 realisations in the non-interacting and 
over 100 samples in the interacting case.}\label{pc3d}
\end{figure}

\section{Distribution of curvatures}\label{curve-distribution}

Not only the typical value of the two-particle level curvatures but also their
probability distribution $P(|c|)$\footnote{We discuss now the probability 
distribution of $|c|$ instead of $\ln|c|$ which, while containing the same 
information, is more convenient for a comparison of the data with distributions
known from the literature.} changes as a function of the interaction. Since the
spectral statistics has recently been found to be critical in the band centre 
at $L=L_1$ \cite{tip2}, we now investigate whether a signature of this 
criticality can be observed in the curvature distribution as well. 

In Figure~\ref{pc3d} the dependence of $P(|c|)$ on energy is plotted for the 
disorder $W=0.5$ chosen such that the one-particle localisation length $L_1$ 
in the centre of the band is equal to the system size and thus only the states 
in the band tails are localised. Picture (a) is for the non-interacting 
two-particle case and (b) is for the value $U=1.25$ where the influence of the 
interaction is close to its maximum.  

In the centre and at the edges of the spectrum the typical curvatures are  
smaller than at other energies (see Figure~\ref{cvse}) and the distribution of 
the curvatures exhibits rather large weights for very small curvatures there. 
This is most pronounced in the non-interacting case and reduced with 
interaction. More interestingly, also the shape of the probability distribution
changes. This is shown in Figure~\ref{pc}, where the probability distributions 
$P(k)$ at different energies are compared without and with interaction. Here, 
we have introduced the normalised curvature $k=c/\overline{|c|}$, with 
$\overline{|c|}$ being the first moment of the distribution $P(|c|)$.   

\subsection{Band centre}

In the non-interacting case, no saturation at small curvatures for the 
distribution in the centre of the band could be found. This indicates that 
strong deviations from a log-normal distribution persist at small curvatures 
which arise due to the mixing of different one-particle energy regimes with 
very different typical curvatures. With interaction the distribution gets 
closer to the form 
\begin{equation}\label{curve-goe}
P_{\rm GOE}(|k|)=\frac{1}{(1+|k|^2)^{3/2}} \, ,
\end{equation}
which was first proposed on the basis of numerical data \cite{zakrzewski} and 
later confirmed analytically for the GOE \cite{oppen-curve}. It was shown that 
(\ref{curve-goe}) holds in the limit $\phi\rightarrow 0$ even when the external
parameter $\phi$ drives the ensemble into another symmetry class 
\cite{braun-mont,fyodorov}. 

\begin{figure}[tb]
\centerline{\epsfxsize=8 cm \epsffile{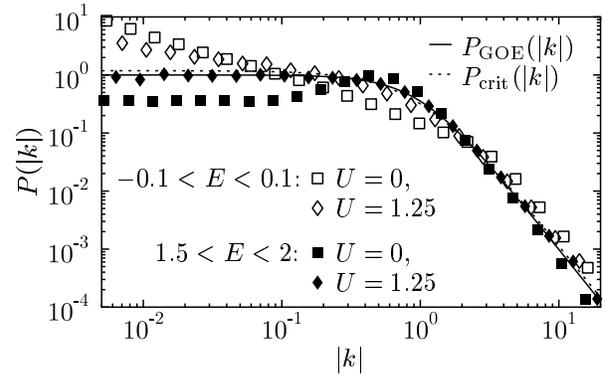}}
\caption{The probability distribution $P(|k|)$ for a ring of size $L=100$, 
averaged over 100 disorder realisations. Open symbols represent the energy 
range in the centre of the spectrum $|E|<0.1$, full symbols stand for 
intermediate positive energy $1.5<E<2$. Squares and diamonds are for data 
points without and with interaction, respectively. The solid and dotted line 
represents the GOE result (\ref{curve-goe}) and the formula for the critical 
distribution (\ref{curve-crit}), respectively.}\label{pc}
\end{figure}

It has been found recently, that the statistics of eigenvalues in the band 
centre at intermediate interaction strength is situated between Poisson and GOE
and indistinguishable from the critical statistics \cite{tip2} which is known 
to hold for several critical systems. Using exactly the same parameters, we 
tried to find a signature of this criticality also in the curvature 
distribution. In Figure~\ref{pc}, we compare our data to the heuristic formula 
\begin{equation}\label{curve-crit}
P_{\rm crit}(|k|)\propto \frac{A}{(1+|k|^\mu)^{3/\mu}}\, ,
\end{equation}
which was found to be very close to numerical results for the Anderson 
transition in three dimensions at the critical point \cite{canali} when 
$\mu\approx 1.58$ was chosen. As can be seen, the difference between the 
proposed $P_{\rm crit}$ (dotted line) and $P_{\rm GOE}$ (solid line) is very 
small and our data for the interacting case in the band centre (open diamonds 
in Figure~\ref{pc}) are inconsistent with both of them. In particular the 
increase at very small curvature values though weakened, persists with 
interaction. Thus, while the spectral statistics for the interacting 
two-particle problem in the band centre \cite{tip2} is indistinguishable from 
the one found at the Anderson transition in 3 dimensions \cite{braun1}, the 
curvature statistics is different from the one found for the Anderson 
transition in reference \cite{canali}. However, the curvature whose properties
are predicted to correspond directly to the spectral statistics is of 
topological origin \cite{ap} and differs qualitatively\footnote{It is assumed 
that the two particles are subject to independent magnetic fluxes $\phi_1$ and 
$\phi_2$. The topological curvatures \cite{ap} of the two-particle energy 
levels $E_j$ are then defined as 
$c_j^{\rm top}=\partial^2 E_j/\partial\phi_1\partial\phi_2
|_{\phi_1=\phi_2=0}$.} from the usual curvature considered in this study. It 
would be interesting to investigate whether the distribution of this 
topological curvature exhibits criticality like the spectral statistics.

\subsection{Finite energy}

In the energy interval $[1.5:2]$ (full symbols in Figure~\ref{pc}), the 
situation is completely different. Without interaction, the distribution 
exhibits a relative minimum at small curvature and a distinct peak close to the
mean curvature, which becomes even more pronounced when the disorder is further
decreased. Such a behaviour is known to occur for the ballistic regime in the 
one-particle curvature statistics \cite{braun}. In one-dimensional systems the 
mean free path is of the order of the localisation length and thus of the order
of the system size for the chosen parameters. Therefore, the one-particle 
dynamics is close to ballistic and the non-interacting two-particle curvatures 
in the given energy range are dominated by the properties of the one-particle 
curvatures. With interaction, the distribution becomes indistinguishable from
the random matrix theory result (\ref{curve-goe}) which is known to be valid in
the diffusive regime of the one-particle problem. This suggests that the
electron-electron scattering makes the originally ballistic dynamics diffusive
and represents a way to understand why the typical curvatures (and the mobility
of the particles) are reduced by interactions in this regime.  

The shape of the probability distribution at the edges of the two-particle 
band, where localised states dominate, is close to the one shown in 
Figure~\ref{pc} for the band centre (open symbols). The interaction drives the 
distribution towards the GOE form, which is a consequence of the mixing of 
one-body states by the interaction. At stronger disorder or larger system size,
eventually all of the states are localised and the probability distribution of 
the two-particle level curvatures is the same in all energy ranges.

\section{Conclusions}

We have studied numerically the two-particle level curvatures in disordered 
chains as a function of the strength of an on-site interaction between the 
particles. This paper contains two main results: Firstly, we found that the 
typical value of the curvatures is increased by the interactions in the 
localised and reduced in the metallic regime, respectively. In the localised 
case, this is a consequence of Shepelyansky's delocalisation effect \cite{shep}
while in the metallic case electron-electron scattering reduces the mobility of
the particles. Our observation represents the first direct confirmation of a 
recent analytical prediction \cite{ap}. Secondly, we have presented evidence 
for a one-parameter scaling of the interaction-induced change of the curvatures
with a non-interacting conductance. The critical curvature where the data are 
not affected by the interaction corresponds to $L/L_1\approx 6$. While the 
precise relation between many-particle level curvatures and the conductance 
should be studied in detail, this suggests that the conductance of metals is 
reduced by the influence of electron-electron interactions, while the 
conductance of insulators (which is exponentially suppressed by the disorder), 
may be enhanced by the interactions. The same conclusion is reached in a study
of the dynamics of two-particle wave-packets \cite{tip3}.

In addition, we have also found the signature of the duality transformation 
between small and very strong interaction \cite{tip2} in the curvatures and 
investigated the probability distribution of the two-particle level curvatures 
in detail. As expected, the distributions are always driven towards the random 
matrix theory result by the interactions, but starting from different 
non-interacting distributions depending on the energy range. This further 
supports that the different consequences of the presence of interactions in 
metallic and localised regimes, respectively, may be relevant for the 
conductance of real systems. 

The level curvatures of the interacting two-particle system yield relevant 
informations like the change of sign of the interaction effect which cannot be 
obtained directly from studies of the spectral statistics. This is in contrast 
to the one-particle problem where curvature and spectral statistics are closely
related.

\begin{acknowledgement}
We acknowledge stimulating discussions with Gert-Ludwig Ingold, Jean-Louis 
Pichard and Xavier Waintal. The numerical calculations were partly done on the 
IBM SP2 at the Leibniz-Rechenzentrum M\"unchen.
\end{acknowledgement}

\end{document}